\documentclass[preprint]{aastex}

\shorttitle{Cloud Ablation in NGC~4402}
\shortauthors{Crowl et al.} 

\begin{document}

\title{Dense Cloud Ablation and Ram Pressure Stripping of the Virgo
  Spiral NGC~4402}
\author{Hugh H. Crowl and Jeffrey D.P. Kenney}
\affil{Astronomy Department, Yale University, P.O. Box 208101, New
  Haven, CT 06520}
\email{hugh@astro.yale.edu~~kenney@astro.yale.edu}

\author{J.H. van Gorkom}
\affil{ Department of Astronomy, Columbia University, 538 West 120th
  Street, New York, N.Y. 10027}
\email{jvangork@astro.columbia.edu}

\author{Bernd Vollmer}
\affil{CDS, Observatoire Astronomique de Strasbourg, UMR 7550, 11 Rue 
de l'Universite, 67000, France}
\email{bvollmer@newb6.u-strasbg.fr}

\begin{abstract}

We present optical, HI and radio continuum observations of the highly
inclined Virgo Cluster Sc galaxy NGC~4402, which show evidence for
ram-pressure stripping and dense cloud ablation. VLA HI and radio
continuum maps show a truncated gas disk and emission to the northwest
of the main disk emission. In particular, the radio continuum emission
is asymmetrically extended to the north and skewed to the west. The
H$\alpha$ image shows numerous HII complexes along the southern edge
of the gas disk, possibly indicating star formation triggered by the
ICM pressure. BVR images at $0\farcs5$ resolution obtained with the
WIYN Tip-Tilt Imager show a remarkable dust lane morphology: at
half the optical radius, the dust lane of the galaxy curves up and out
of the disk, matching the HI morphology. Large dust plumes extend
upward for $\sim 1.5~\textrm{kpc}$ from luminous young star clusters
at the SE edge of the truncated gas disk. These star clusters are very
blue, indicating very little dust reddening, which suggests dust blown
away by an ICM wind at the leading edge of the interaction. To the
south of the main ridge of interstellar material, where the galaxy is
relatively clean of gas and dust, we have discovered 1 kpc long linear dust
filaments with a position angle that matches the extraplanar radio
continuum tail; we interpret this angle as the projected ICM wind
direction. One of the observed dust filaments has an HII region at its
head. We interpret these dust filaments as large, dense clouds which
were initially left behind as the low-density ISM is stripped, but are
then ablated by the ICM wind. These results provide striking new evidence
on the fate of molecular clouds in stripped cluster galaxies.

\end{abstract}


\section{Introduction}

The morphology-density relationship \citep{oemler74,ms77,dressler97} is
one of the clearest examples of the effect that clusters have on their
member galaxies. There are several cluster processes that may
contribute to this observed effect \citep{treu03}. Mergers, harassment
(repeated high speed galaxy-galaxy interactions), and tidal truncation
of the outer galactic regions by the cluster potential
\citep{merritt84,natarajan02} are among the suggested gravitational
effects for galactic morphological transformation in clusterseffects of the intracluster medium (ICM) on the galactic interstellar
medium (ISM) include ram pressure stripping \citep{gunn72,abadi99},
turbulent and viscous stripping \citep{nulsen82,toniazzo01}, and the
thermal evaporation of the ISM by the hot ICM \citep{cowie77}. It is
believed that ISM-ICM stripping (which includes both ram-pressure
stripping and turbulent viscous stripping) may be among the most
important processes in the transformation of late-type cluster spirals
into Sa's and S0's. This process, which removes gas from the galaxy
but leaves the stars unperturbed, results in galaxies with truncated
gas and star-forming disks
\citep{warmels88,cayatte90,koop04a,koop04b}. Simulations of ISM-ICM
interactions (e.g. \citealp{quilis00}; \citealp{ss01};
\citealp{vollmer01}) have shown that a smooth, uniform ISM can be
stripped from the outer regions of galaxies in clusters via ram
pressure. However, such simulations necessarily assume an overly
simplistic ISM and, hence, are unable to model the multi-phase ISM as
it exists in spiral galaxies. In particular, the fate of star-forming
molecular clouds must be known to fully understand how stripping
affects cluster galaxy evolution. Studying the details of an ISM-ICM
interaction can help us learn what role gas stripping plays in the
morphology-density relation. There is abundant evidence that NGC 4402
is undergoing such an interaction.

NGC~4402 is an HI-deficient, $0.3 L_*$ Sc galaxy located near the
center of the Virgo cluster, $1 \fdg 4$ (390 kpc projected) from the
giant elliptical M87. It is also the nearest spiral galaxy to the
giant elliptical M86, making it plausible that this galaxy is
associated with the M86 subcluster \citep{schindler99}. NGC~4402 is
highly inclined ($i = 80 \pm 3\degr$), making it possible to
distinguish stripped extraplanar gas from disk gas. Its line-of-sight
velocity of $237~\textrm{km/s}$ \citep{rubin99} means that it is
blueshifted relative to the mean cluster velocity by nearly $800~
\textrm{km/s}$. This, in turn, means that the near side of the galaxy
contains the leading edge of the ISM-ICM interaction, allowing us to
observe key details of the ISM-ICM stripping in this galaxy. In
this paper, we present new high-resolution (0\farcs5) optical images,
along with 21cm VLA radio maps of NGC~4402. We show that both the
radio and optical data provide a strong constraint on the ICM wind
direction and the optical data suggest dense cloud ablation by the ICM
wind. Although present millimeter interferometers do not have the sensitivity or
resolution to show what happens to giant molecular clouds in stripped
galaxies, the high-resolution optical images of star formation and dust
extinction suggest what may happen to large dense gas clouds in NGC~4402. 

\section{Observations}

\subsection{Radio Data}

NGC~4402 was observed with the VLA\footnotemark{} in C array in January of 2003 as
part of a study of four edge-on spiral galaxies in the Virgo cluster. We used the 21-cm
spectral line mode with online Hanning smoothing and the system tuned
to a central velocity of 200 km s$^{-1}$. In this configuration, the
system has a bandwidth of 3.125 MHz and 63 channels, each with a width
of $10.4$ km s$^{-1}$. The data were calibrated, mapped and ``CLEANED'' in AIPS using standard
techniques and procedures. Maps were made with the IMAGR task with a
ROBUST weighting parameter of +1, resulting in a beam size of $17\farcs4
\times 15\farcs2$ ($\sim 1.2 \textrm{~kpc}$, assuming a distance of 16
Mpc). Thirty of the channels show HI
emission at the $>3\sigma$ level (where $\sigma = 0.32 \textrm{mJy/beam}$). These channels were combined to make moment maps of
the HI emission. Fifteen line-free channels were combined to make the
continuum map. Maps of the total HI emission and radio
continuum emission are shown in Figure 1. We detect a total HI flux of
$6.50 \pm 0.55$ Jy km/s. This compares
with an average single dish HI flux in the literature of $7.16\pm0.70$
Jy km/s \citep{giovanelli83,helou84,giovanardi85}, suggesting that there is no significant flux missing in our
interferometer observations. The HI mass of this galaxy as measured
from our data is $3.9 \times 10^8 M_\odot$. This galaxy is moderately HI deficient, with a deficiency
parameter of
0.61 \citep{giovanelli83}, roughly 25\% of the ``normal'' HI content for a
galaxy in a low-density environment.

\footnotetext{The VLA is operated by the National Radio Astronomy Observatory
(NRAO) which is a facility of the National Science Foundation, 
operated under cooperative agreement by Associated Universities, Inc.}

\subsection{Optical Imaging Data}

The optical data were taken during two separate observing runs at
the WIYN telescope. The high-resolution
imaging data were taken over two nights in March of 2004 using the
WIYN Tip-Tilt Module (WTTM) on the WIYN 3.5m telescope. WTTM is a
first-order adaptive optics system that allows us to correct the
already excellent seeing at WIYN by $0\farcs1$ to $0\farcs15$ through
use of a bright guide star \citep{claver03}. As a result we were able to obtain $B$,
$V$, and $R$ images with $0\farcs5$ seeing. The deep $R$ and narrowband $H\alpha$ images
were taken in April of 2000 at WIYN using the MiniMosaic imager. The
resulting images have an average seeing of approximately
$1\arcsec$. For both data sets, standard IRAF image reduction tasks were used to
bias correct, flat field and combine the images. The $R$-band image was used to
subtract the continuum emission from the $H \alpha$ narrowband image
and produce an $H \alpha + \rm{[NII]}$ image. The MiniMosaic $R$ and
$H\alpha$ images are shown in greyscale in Figures 1 \& 2a, together
with the HI and radio continuum contour maps. A higher resolution WTTM $B-R$ image
is shown in Figure 2b and a color version of the higher resolution BVR
image is shown in Figure 4.

\section{Results}

The stellar disk appears undisturbed, since outer isophotes of the
R-band image are elliptical, concentric, and at a constant
position angle. Within the disk, both the HI and radio
continuum (Figures 1a and 1b) are truncated (i.e. no emission at a $2 \sigma$ level) at
$0.6-0.7~R_{25}$ ($= 70\arcsec-82\arcsec = 5.4-6.4~\textrm{kpc}$), demonstrating that the undisturbed stellar disk extends well beyond 
the location of the gas. Throughout the part of the disk still
containing HI and radio continuum emission, heavy dust lanes are evident (Figure
2b).

The truncation radius is similar for the HI and radio continuum on the eastern edge
of the galaxy ($\sim 0.6~R_{25}$), but is more
complicated in the West. If one considers only the symmetric disk components, the
radio continuum and HI emission both extend to $\sim 0.7~R_{25}$ along the major
axis in the West. However, there is
a component of the radio continuum halo north of the major axis that extends
to $0.9~R_{25}$, and a significant component of the HI emission north
of the major axis that
extends to $1.1~R_{25}$. The apparent extraplanar emission in both the
HI and radio continuum maps is displaced to the northwest of the disk emission, 
consistent with an ICM wind acting from the SE direction,
although the distributions of extraplanar emission differ. While there are many processes that may drive galactic
evolution in clusters, ICM-ISM stripping is the
only one that can asymmetrically strip the gas from the outer parts of a galaxy
without disturbing the stars.

The radio continuum emission is asymmetrically extended away from the
disk midplane.
South of the major axis, the radio continuum contours are compressed, consistent
with a marginally resolved or unresolved distribution. In contrast, north of the major axis, the radio continuum contours are stretched out,
reaching at least $60 \arcsec$ (500 pc) to the north of the
disk. North of the major axis, there is a 50\% excess of radio continuum emission compared
to the emission from south of the major axis. Many edge-on galaxies are observed to have extended radio continuum halos,
both inside and outside of clusters \citep{irwin99}. The causes of these extended halos are not
well understood, but are likely related to the star forming activity
in the disk. Since 25\% of the edge-on galaxies in the \citet{irwin99}
sample are observed to have larger radio continuum extents that NGC~4402, we cannot say with certainty that the large extent of the
northern radio continuum halo is due to the ICM wind. However, what is clearly unusual
about the radio continuum halo in NGC~4402 is the large asymmetry on the two sides
of the major axis and a clear skewing of the northern radio continuum halo toward
the NW. Figures 1a \& 3 show that the 2 sides of the radio continuum tail form fairly sharp
ridges with 
well-determined position angles of $-49^\circ$ (east) and $-47^\circ$
(west). 
These relatively sharp boundaries may be indicative of the
projected ICM wind direction. No other edge-on galaxy in the \citet{irwin99} sample show
such an asymmetry or skew.

The HI map (Figure 1b) shows 2 possible extraplanar features,
a small one $50 \arcsec$ (3.9 kpc) W and $40 \arcsec$ (3.1 kpc) N of
the nucleus,
apparently lying above the main disk HI emission, 
and a larger curved feature extending for $0\farcm8$ beyond the Western 
disk truncation radius, lying $25 \arcsec$ N of the major
axis. The velocity structure of these components can be seen in the Moment 1
map (Figure 1c). The gas to the north of the disk has a higher line of sight
velocity (and lower galactocentric velocity) by $\sim
20\textrm{km/s}$ than the corresponding disk gas along the major
axis. We can put an upper limit on the
amount of extraplanar HI by assuming the disk emission is symmetric about
the major axis, and subtracting a disk component from regions north of
the major axis. Approximately 6\% of the total HI flux (0.40 Jy km/s) is outside
of the symmetric disk, corresponding to $2.4 \times
10^7 M_\odot$. This ``extraplanar'' HI fraction and mass are much lower than 
the 40\% and 1.5$ \times 10^8 M_\odot$ detected in the highly inclined
Virgo spiral NGC~4522 \citep{kenney04}, a good candidate for
ongoing stripping. The 6\% excess of HI flux is also significantly lower than the 50\%
excess of radio continuum emission north of the major axis in NGC~4402.

The total H$\alpha$ luminosity of the galaxy is $7.9 \times
10^{40} \rm{erg/s}$, which corresponds to a SFR of $0.6 \rm{M_\odot/yr}$, whereas the FIR luminosity
from IRAS fluxes \citep{soifer89} corresponds to a SFR of $1.2 \rm{M_\odot/yr}$. 
The FIR-based SFR is unsurprisingly higher for this dusty, highly-inclined 
galaxy, but the H$\alpha$ image still shows a large fraction of the
star formation activity in the galaxy.
The $H\alpha$ emission (Figure 2a) is strong along the southern edge of the galaxy. Asymmetric H$\alpha$
enhancements at the gas truncation radius have been observed in other
galaxies in Virgo and other clusters \citep{koop04b, vogt04}, and
interpreted as a sign of ongoing pressure. H$\alpha$ emission from
star-forming complexes is strong in both the SE and SW, but the
complexes in the SE are much bluer in the $B-R$ and color images
(Figures 2b \& 3). It appears that the
SE star clusters suffer less dust reddening and extinction because the
ICM wind has pushed away much of the dust at the leading edge of the
galaxy, an interpretation supported by the dust plumes extending
upward from the SE star forming regions. This provides evidence that
the SE region, opposite the NW radio continuum tail, is the leading edge of the interaction.

Our H$\alpha$ image (Figure 2a) also shows the recently-discovered
extraplanar HII region northwest of the main disk \citep{cortese04},
which likely formed from stripped gas.
This HII region is located within the extraplanar radio continuum tail, and
close to, but not coincident with the ``extraplanar'' HI.
Approximately 0.3\% of the total  H$\alpha$ luminosity is from this
extraplanar HII region, much lower than the extraplanar  H$\alpha$ fraction of
10\% in NGC 4522 \citep{kenney04}.

The HI and radio continuum distributions are different, indicating that they trace different
components of the ISM. HI emission traces the warm, neutral medium and
non-thermal radio continuum emission traces relativistic electrons spiraling in
magnetic fields, perhaps produced by supernovae resulting from
recent star formation. The HI is more extended than the radio continuum along the
major axis in the west. This could simply be the result
of the low radio continuum to HI ratio generally observed in
the outer disks of spirals, due to the low
efficiency of outer disk HI in forming stars \citep{kennicutt89}.
Consequently, there are few products of star formation, 
including radio continuum emission, in outer disk gas. Figure 2a shows that the radio continuum
emission in the disk is co-extensive with regions of star formation
along the major axis,
and that the westernmost HI has no associated HII
regions. In the north, the radio continuum emission is more extended than the HI
along the minor axis, as discussed below.

The high-resolution images of the dust lanes seen in the $B-R$ image (Figure 2b) and BVR color image (Figure 4)
are especially striking. The strongest dust lanes are across the southern
part of the stellar disk, indicating that the southern side of the
galaxy is the near side. The color image (Figure 4) shows that the galaxy starlight 
south of the main dust band is relatively blue, while, in the north, it 
is significantly reddened by the displaced dust. This is
\emph{opposite} what is normally observed in undisturbed galaxies, where the near side of the galaxy
appears to be more reddened. The $BVR$ and $B-R$ images show a fairly well-defined boundary between the region of NGC~4402 that still contains large amounts of dust,
and the region that has been swept mostly clean. There are very few
dust extinction features observed south of the main boundary. While dust is an
excellent ISM tracer, we can only observe it via optical extinction if the dust is
in front of most of the stars. Therefore, the lack of dust extinction
features south of the main truncation boundary indicate that there is
very little dust above or near the midplane of the outer southern stellar disk. Heavy
dust lanes are evident throughout that part of the disk still
containing HI and radio continuum emission. In both the east and the west, at the
HI truncation radii, the dust lanes curve up and away from the plane
of the stellar disk. The departure from the disk is more gradual in
the west, and more pronounced in the east, similar to the radio continuum and HI
distributions. In the East/Southeast, which we think may be the leading edge of the gas
disk, the dust features curve upward from the luminous star forming
complexes for at least $20 \arcsec$=1.5 kpc. The $B-R$ map (Figure 2b)
shows that the HI contours appear to cut off \emph{inside} the dust
distribution, suggesting that the less dense gas in this part of the
galaxy has already been stripped.

\section{Discussion}

The sharpness of the southern boundary in the three color image
(Figure 4) reveals important information
on how the multi-phase ISM reacts to ICM pressure. In the HI map (Figure 1b), the southern boundary is not
smooth, but featured, perhaps due to ISM structure within the
disk. The optical image shows that the region south of the main ISM boundary
contains a few faint dust features
and young blue star clusters. The most remarkable dust features south of this boundary are 2 linear
dust filaments, shown in Figure 4, 
with widths of $\sim$1-2'' (80-150 pc), and lengths of $7 \arcsec$ (0.5 kpc) 
and $13\arcsec$ (1.0 kpc), respectively. These 2 filaments have nearly
the same position angle ($-40^\circ$ in the east and
$-36^\circ$ in the west) and point roughly in the same direction as the
radio continuum tail, indicating that their morphology is associated with the
ICM wind direction. The eastern filament shows two pieces of additional evidence of
dense gas. First, there is a star formation region that appears to be
associated with the head of the filament. Near the eastern dust
filament, there are several
($\sim 10$) additional blue star clusters that are also south of the main
truncation edge. Secondly, there is a weak feature
in the HI map possibly associated with that filament. We can
estimate lower limits on the gas column densities, $N_H$, and gas
masses of these filaments using the relative extinction of dust
features \citep{howk97}. The eastern dust filament has an average relative
extinction of $a_V = 0.27$ over an area of 6.5 arcsec$^2$ ($4.0 \times
10^4$ pc$^2$) and the
western filament has an average relative extinction of $a_V= 0.13$
over an area of 13 arcsec$^2$ ($8.1 \times 10^4$ pc$^2$). Assuming a ratio of total to selective absorption of $R_V=3.6$, and using the method of \citet{howk97}, we
find both the eastern and western dust filaments have $M_{gas} > 2 \times 10^5
M_\odot$. This is a strict
lower limit, since foreground stars (which are numerous if the
filament head is near the disk midplane) and small dense lumps in the
filaments can cause us to underestimate the extinction, making the
true cloud mass higher.

Extraplanar dust filaments have been observed in many galaxies both
inside and outside of galaxy clusters \citep{howk99, alton00}. In many
cases, these filaments are associated with fountains of gas and dust
driven by star formation in the disk. In
NGC~4402, there are several different types of dust features
observed. The most prominent, perhaps, is the large plume of dust
(with dimensions $15 \arcsec$ x $6\arcsec = 1.3~\rm{kpc}$ x $0.5~\rm{kpc}$) and other dusty clumps
north of the minimally reddened star clusters in the East. Once again
using the method shown in \citet{howk97}, we can put lower limits on
the mass of the two largest dust plumes (outlined in the close-up image of
Figure 4). The large plume furthest to the East has a mass of $M > 2 \times
10^6 M_\odot$ and the plume $7\arcsec$ west of it has a mass of $M > 6
\times 10^5 M_\odot$. West of the eastern
dust plumes and north of the main dust band, we
observe a complex filamentary structure similar to those observed in
other galaxies. These
structures are similar in morphology to the ``Irregular Clouds'' and
``Arcs'' observed in NGC 891 \citep{howk00}, and are likely due to
star formation activity in the disk. Finally, $1\farcm5$
west of the minor axis and beyond, there are long, curved, smooth dust
features. The lack of strong
star formation at this location has apparently left these dust
features undisturbed. The filaments observed south of the disk in
NGC~4402 (described above), which we believe are \emph{not} caused by star formation, are different in several ways. First,
the two large filaments that we have observed are linear. By contrast, filaments observed in other
galaxies (i.e. NGC 891, NGC 3628) and north of the disk of NGC~4402 have more complex morphologies with
significant substructure and curvature, and a large variety of position angles. This is the morphology that one would
expect of dust ejected out from different star forming regions and falling
back to the disk via gravity. Secondly, there is very little dust
observed in the
region of the filaments south of the boundary,
suggesting that the area has been swept relatively clean of the less-dense
dust. Finally, the filaments are aligned with each other, and with the
radio continuum tail.

The most straightforward interpretation of the observed southern filaments is
that they are dense clouds in the disk midplane being ablated by the
hot ICM wind. We propose that the southern tips of these filaments are \emph{not}
extraplanar, but are the remnants of dense clouds in the
disk, which has otherwise been swept mostly clear of its ISM. If this is
true, then the ``tails'' of these filaments are extraplanar, swept up by the ICM
wind. This suggests that some clouds,
probably the biggest and densest giant molecular clouds,
do not get stripped out together with the rest of the lower density ISM,
but are left behind, at least initially. Although we do not have
direct evidence of molecular gas, the sizes (80-150 pc) and masses
($> 2\times 10^5 M_\odot$) of these clouds are consistent with the largest giant molecular clouds in
our galaxy \citep{dame86} or the cores of Giant Molecular Associations
(GMAs) (with typical sizes of $\sim 300$ pc \citep{vogel88}). In a typical spiral, over a small
galactocentric annulus, there are very few of these largest clouds. If these are the remnants
of GMAs, it is not surprising that
they survived longer than any other ISM tracer. If the observed dust
clumps are GMAs/GMCs being ablated,
one would expect to see head-tail features from ablation with a
variety of sizes, reflecting the GMA/GMC mass spectrum. With our
observations at $0\farcs5 = 40~\textrm{pc}$ resolution, we can only
see the most massive features. Higher resolution observations may show
a continuum of structure from south to north across the disk: smaller
molecular clouds may be visible at smaller galactic radii. The blue star clusters in the SE are perhaps the remnants of
these other molecular cloud cores. Calculations show
that molecular clouds are too dense by at least a factor of $\sim$10
to be directly stripped by ram pressure in a cluster like Virgo
\citep{kenney89,kenney04}. Our observations of NGC~4402 suggest that while some clouds may survive the initial
stripping, they are eventually destroyed
by the ICM wind, probably aided by star formation and normal cloud
evolution, which act themselves on timescales of $\sim 10^7$ yrs
\citep{larson03} to destroy molecular clouds. The cloud is ablated in an
outside-in manner: the outer, less-dense, parts of the clump of
material are stripped off first. This material then streams behind the
clump to form a head-tail morphology. In the eastern filament, this
wind has either triggered star formation, or ablation has exposed an
already-existing star-forming region. At the leading edge of the disk,
molecular clouds can probably be triggered to form stars by ISM pressure. While
a fraction of the cloud mass is converted into newly formed stars,
most will be returned to the lower density ISM by the destructive
energy of the young stars \citep{larson03}, where it can be more
easily stripped.

An important parameter in systems such as NGC~4402 is the stage of the
ISM-ICM interaction. In the case of NGC~4402, the optical image provides compelling evidence of an ongoing
interaction. Specifically, the upward swirl of dust from minimally
reddened luminous star clusters at the gas truncation radius in the
SE, the aligned linear dust filaments just beyond the gas
truncation radius, and the extended and skewed radio continuum halo all strongly suggest that NGC~4402 is currently
experiencing ICM pressure. This pressure is apparently strong enough
to disturb the radio halo, 
ablate the molecular cloud cores, and push dust up into the halo of the
galaxy. However, simulations of ram
pressure stripping \citep{quilis00,ss01,vollmer01} show significant
amounts of extraplanar gas close to the disk in the early phases of
stripping. Although 75\% of the original HI is missing, there is very little extraplanar HI near the disk of NGC~4402, in contrast to NGC 4522
\citep{kenney04}, where 40\% of the HI is extraplanar. This result
seems at odds with a near-peak ram pressure scenario. The seemingly
conflicting results yield two possible explanations: 1. The
galaxy is in an early phase of stripping, but the stripped extraplanar
gas is no longer in the form of HI. The gas has
become ionized shortly after leaving the disk so that it no longer appears in the HI phase. Indeed, heating and ionization
must happen at some point, since this galaxy is quite HI deficient. If this is correct, we would expect diffuse $H\alpha$ and
X-Ray emission from this heated and ionized gas. In this scenario, the
NW radio continuum tail may be tracing the general ISM stripped from the disk of
NGC~4402. 2. The galaxy is in a
significantly later phase of stripping. Much of the outer-disk HI has
been pushed far away from the disk, but the galaxy is still experiencing pressure. If the galaxy is in a later phase of stripping, the dearth of HI is expected. In fact, this is in line with the
later stages of simulations of ram pressure \citep{ss01,vollmer01}, which show that after
$\sim$ 100-500 Myrs, the pressure results in a truncated gas disk with little or
no extraplanar gas near the galaxy. Other simulations \citep{quilis00}
have shown that viscous and turbulent stripping \citep{nulsen82} can be important over
larger timescales and, therefore, may become more important than ram
pressure at times following peak pressure. So, while bulk
ram-pressure stripping of the neutral gas may have occurred long ago,
ongoing viscous and turbulent stripping may continue to strip gas from
the edge of the disk
well after peak pressure. In this scenario, the large
vertical extent of the northern radio continuum emission does not arise from the
stripping of disk gas, but is a predominantly normal radio continuum halo
(i.e. \citealp{irwin99}) shaped by the ICM wind, which
compresses the halo in the south-east and extends it to the
north-west, giving the halo its asymmetry and skew. 

To determine if our interpretations are realistic, we have made
order-of-magnitude calculations of the internal pressure in the
galaxy and the pressure from the ICM wind. If the radio continuum halo can be
shaped by the ICM wind, the ICM
wind pressure must be at least as large as the cosmic ray energy density, as
cosmic rays are the
source of the radio continuum halo. Assuming a spectral index of $\alpha =
-0.86$ \citep{vollmer04} and measuring the radio continuum surface brightness in a
small region near the truncation radius, we estimate the cosmic ray
energy density to be $u_{min}^{CR} \gtrsim 1 \times 10^{-12}$ ergs
cm$^{-3}$, using the method outlined in \citet{irwin99}. This is a
lower limit of the cosmic ray energy density, as the effective
emitting volume is likely smaller than the total volume along
the line of sight. The strength
of the ICM wind depends on the density of the ICM and the
relative velocity of the ICM and NGC~4402. From X-Ray surface brightness measurements
\citep{schindler99}, the ICM density is estimated to be $\rho_{ICM} =
5 \times 10^{-12}~\rm{g~cm^{-3}}$ at a cluster radius of $r=0.4$
Mpc. With this density and a typical galaxy velocity of 1300 km/s
\citep{binggeli93}, the ICM ram pressure would be $\rho v^2 = P_{ICM} \sim
7 \times 10^{-12}$~dynes cm$^{-2}$. The true ram pressure could be
higher or lower than this. The ICM density and, therefore, the ram pressure could be lower, since NGC~4402's
clustercentric distance is probably higher than its projected
distance. In contrast, the ram pressure could be higher if the ICM has
bulk motions \citep{dupke01} and the motion of the galaxy with respect
to the ICM is much
higher than galaxy velocity dispersions indicate
\citep{kenney04}. Even with these uncertainties, it seems reasonable
that the present ram pressure is strong enough to disturb the radio continuum halo.

The ICM pressure needed to strip
neutral gas from the disk is significantly larger than that needed to
disturb the radio continuum halo. The pressure needed to strip neutral gas can be estimated by 
considering the gravitational restoring force for a region of the
galaxy (see NGC~4522, \citealp{kenney04}). Since NGC~4522 and NGC~4402
are similar in luminosity and $v_{max}$, it is reasonable to compare them. However, since the
gas truncation radius in NGC~4402 (0.6-0.7 $R_{25} \approx 5$ kpc) is larger than
that of NGC~4522 (0.4 $R_{25} \approx 3$ kpc), the R-band surface
brightness and the disk mass surface density at the truncation radius
is lower by a factor of two. For a gas
surface density of $\sigma_{gas} = 10 M_\odot$ pc$^{-2}$ and an
encounter angle of $45 \degr$, the gravitational restoring force per
unit area is $\sigma_{gas} \frac{d\phi}{dz} \sim 2 \times 10^{-11}$ dynes
cm$^{-2}$. This is larger by a factor of 3 than the above estimate
of the ICM ram pressure for a static ICM and larger by a factor of 20
than the pressure needed to disturb the radio continuum
halo. As mentioned above, turbulent and viscous stripping may be an
important effect that has not been considered in these simple
calculations. Such an effect should be most important near the edge of
the galactic disk, which is where the largest disturbances of dust are
observed in the optical image (Figure 4).

Gas infall can occur, either with large amounts of gas after peak pressure
\citep{vollmer01}, or with a small amount of gas in a
constant ram pressure scenario \citep{ss01}. Thus, we have considered whether the observations, particularly the
linear dust filaments, might be due to gas infalling after peak
pressure. In scenarios such as these, one could, in principle, produce head-tail gas/dust
features. However, an infall scenario does not easily account
for the following observations: 1. The alignment of the filaments with each other and with the radio continuum
tail. If the filaments were infalling, one would expect them to fall
back in the direction of the disk and galactic center and to have
orientations that reflect this. In
particular, the alignment of the filaments on opposite sides of the
galaxy is not consistent with this, and therefore not easily explained by gas fallback. 2. The proximity of the SE linear filament, with the HII
region at its head, to the minimally-reddened SE disk star
forming regions and the blue star clusters. Additionally, an upward
swirl of dust from other star forming complexes is evident at this
location. By contrast, these observations are naturally explained by ongoing ICM
pressure and dense cloud ablation.

These observations of NGC~4402 have furthered our
understanding of ram pressure stripping in clusters, and have allowed
us to question more deeply the nature of these processes. We have learned
that: \textbf{1. Not all dense clouds are strongly coupled to the lower-density ISM,
  and it seems possible to strip low-density gas without stripping the dense
  clouds. 2. These dense clouds do not survive indefinitely and are
  eventually ablated by the ICM wind.} It is clear that NGC~4402 is
currently experiencing a wind pressure strong enough to distort the radio continuum
halo and ablate the dense clouds in the disk. However, our observations raise interesting questions on the
timescales over which these processes happen. Specifically:
\textbf{1. Is
  active stripping of the neutral gas in the disk occurring now? 2. When was the outer
  disk of NGC~4402 stripped? 3. On what timescales are molecular
  clouds destroyed by ablation?} Further observations to constrain the ages of stars in the outer stellar disk
may allow us to better determine the stripping history of this
galaxy. Despite timescale uncertainties, it is clear that molecular clouds do not prevent the
entire ISM of a galaxy from being stripped. This makes it feasible
that the cessation of star formation and subsequent disk fading in
cluster spirals stripped by ram pressure contributes to the
transformation of these galaxies into early type spirals and
lenticulars.

We gratefully acknowledge Richard Rand for helpful discussions, Gene
McDougal and the rest of the WIYN staff for effective and efficient observing support, and Mark
Hanna of NOAO for assistance in the creation of a color image. We also
gratefully acknowledge the referee, Curt Struck, for helpful
comments which improved the paper. This is research supported by NSF
grants AST-0071251 and AST-0098294.

\clearpage

\begin{figure}
\begin{center}
\includegraphics[width=5in]{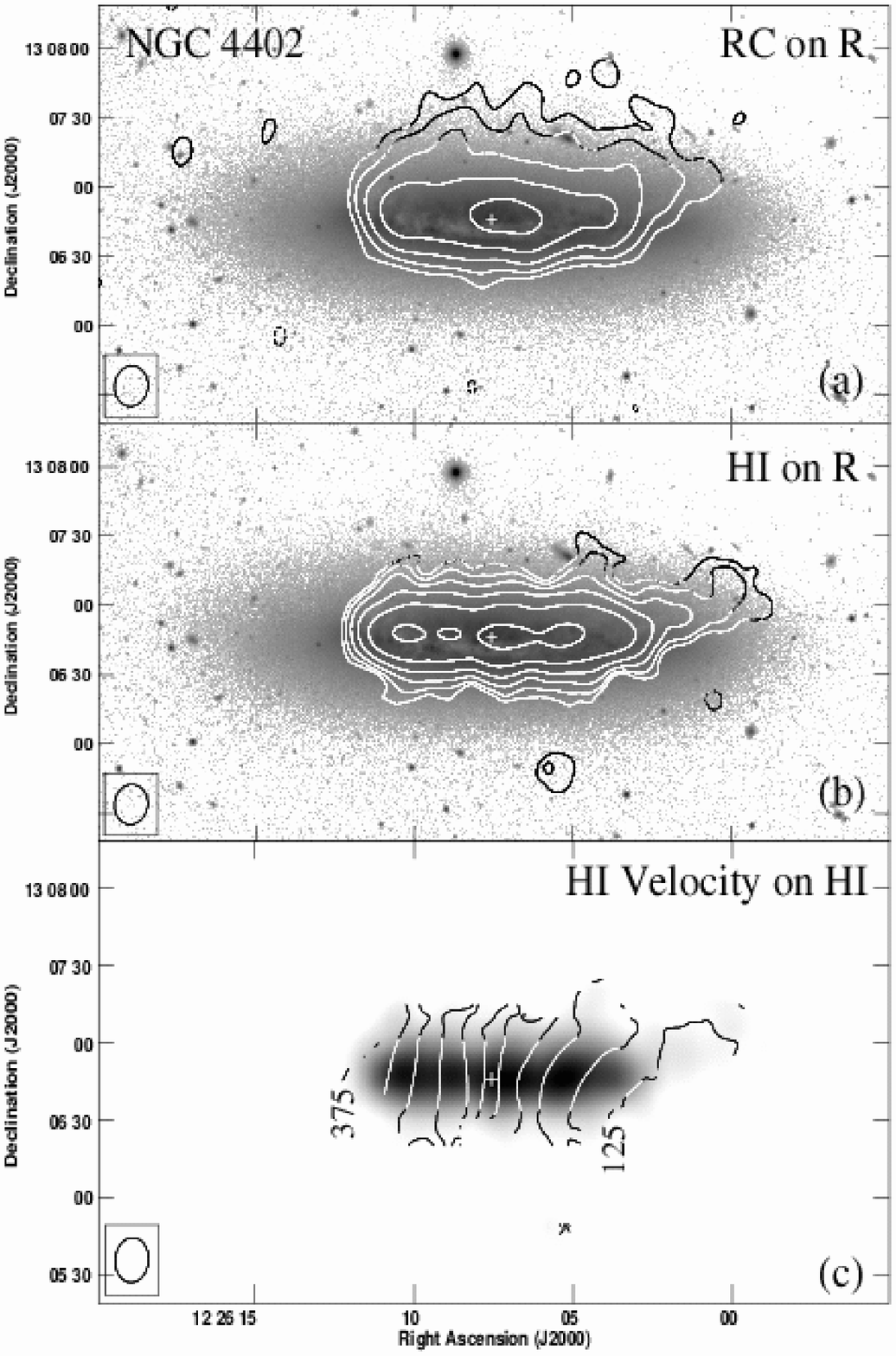}
\end{center}
\caption{(a) 1.4 GHz Radio Continuum contours overlayed on a MiniMosaic $R$-band
  image. The contours are $-0.46~\rm{mJy/beam}$ and then
  range from $0.46~\rm{mJy/beam}$ to $14~\rm{mJy/beam}$ by factors of two. The $R$ image is shown in a
  logarithmic stretch. $R_{25}$ is located approximately 1.7 times
  further out than the lowest contour level. (b) HI contours on the
  $R$-band image. Contours range from $0.6~\rm{M_\odot~pc^{-2}} = 8
  \times 10^{19} \rm{cm^{-2}}$ to $19.2~\rm{M_\odot~pc^{-2}} = 2.6 \times 10^{21} \rm{cm^{-2}}$ by factors of two. (c) HI velocity contours on
  HI grayscale. Contours range from 125 km/s in the west to 375 km/s
  in the east and are incremented by 25 km/s.}
\end{figure}

\begin{figure}
\begin{center}
\includegraphics[width=6in]{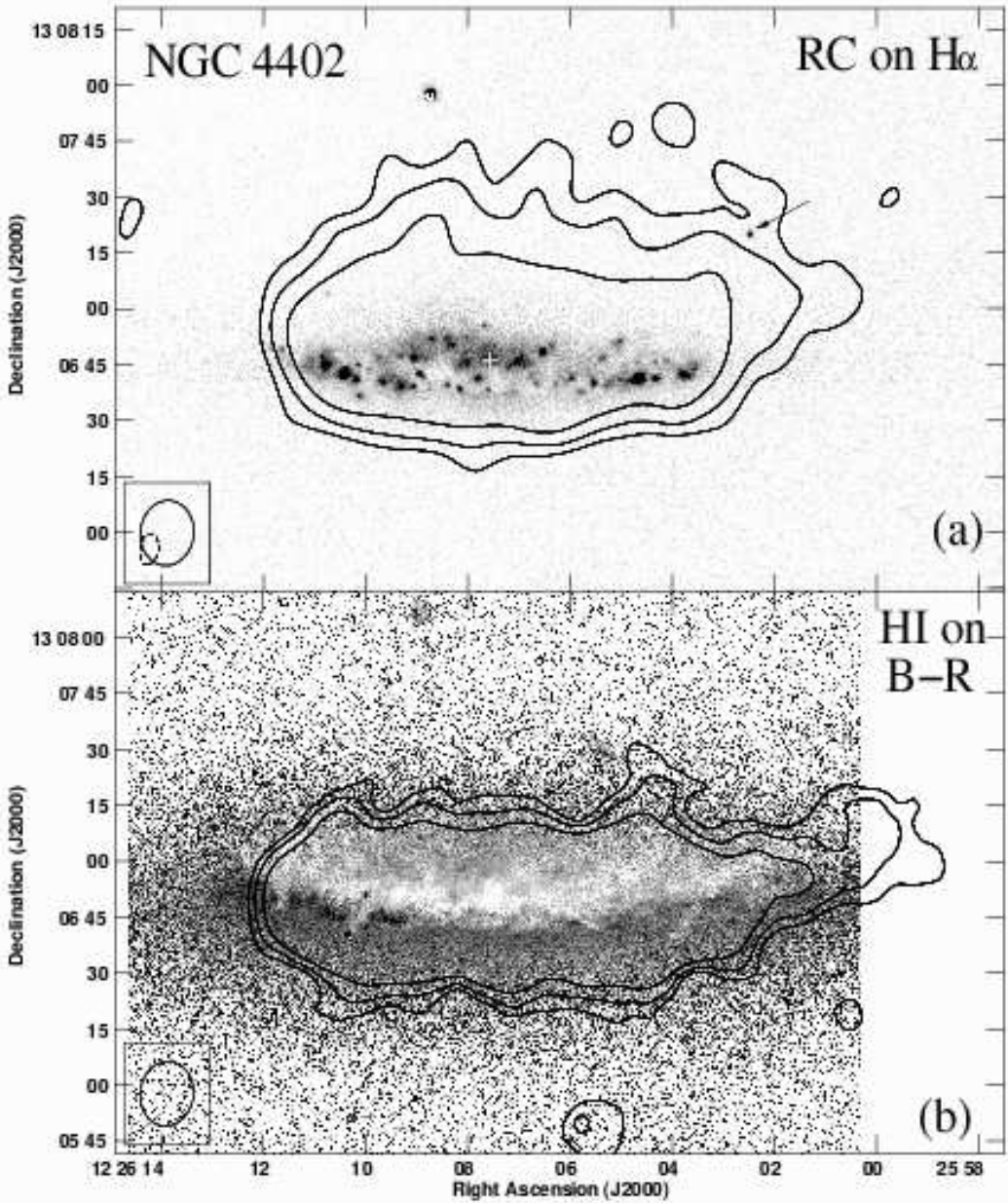}
\end{center}
\caption{(a) $H \alpha$ image showing the outer three radio
  continuum contours. The extraplanar HII region from
  \citet{cortese04} is marked with an arrow. (b) B-R image from WTTM
  images showing dust lane detail in the galaxy, along with the outer
  3 HI contours. Note the very blue star clusters in the SE, probably
  due to minimal reddening by dust.}
\end{figure}

\begin{figure}
\includegraphics[width=6.5in]{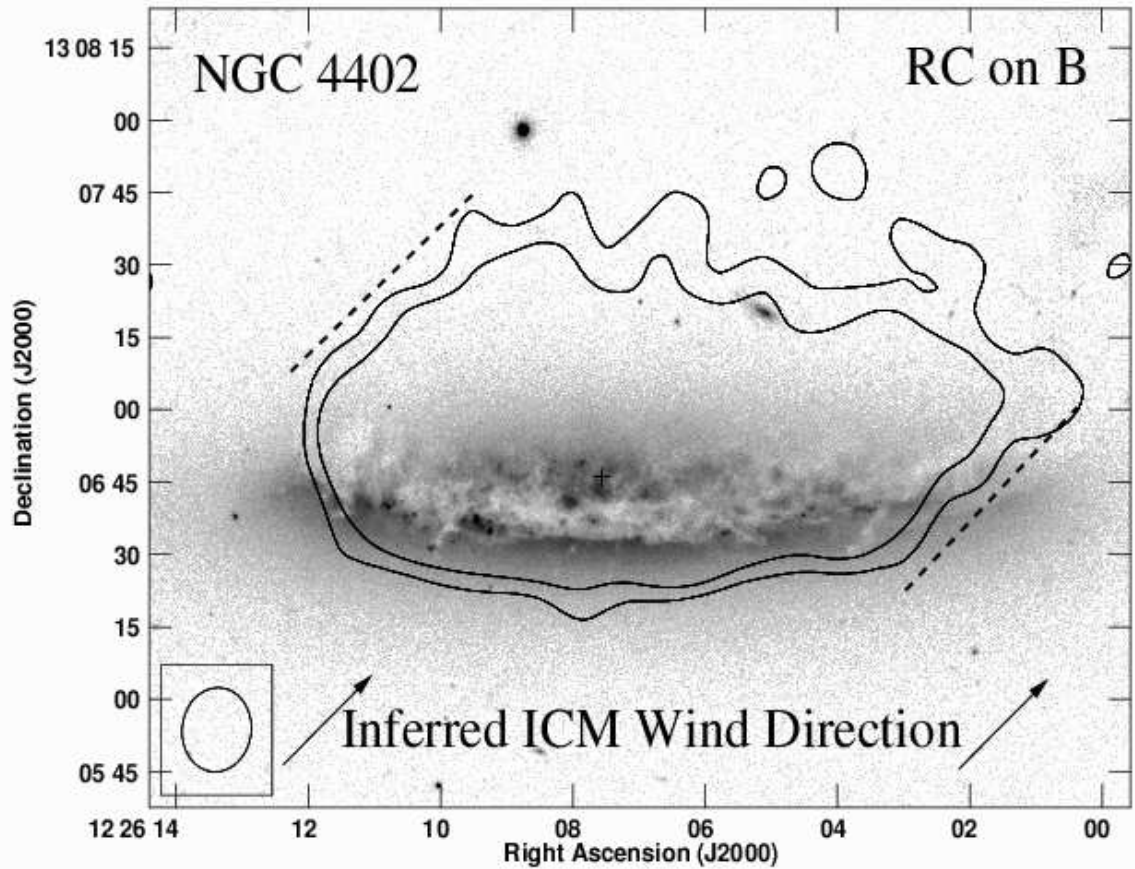}
\caption{Outer two 1.4 GHz radio continuum contours on WTTM B
  image. The dashed lines indicate the ``sharp ridges with well
  defined position angles'' discussed in the text. The inferred
  projected ICM wind direction (PA= $-43\degr$), as calculated from
  the average of the position angles of the filaments and the position
  angles of the radio continuum ridges, is indicated with arrows at
  the bottom of the image.}
\end{figure}

\begin{figure}
\includegraphics[width=6.5in]{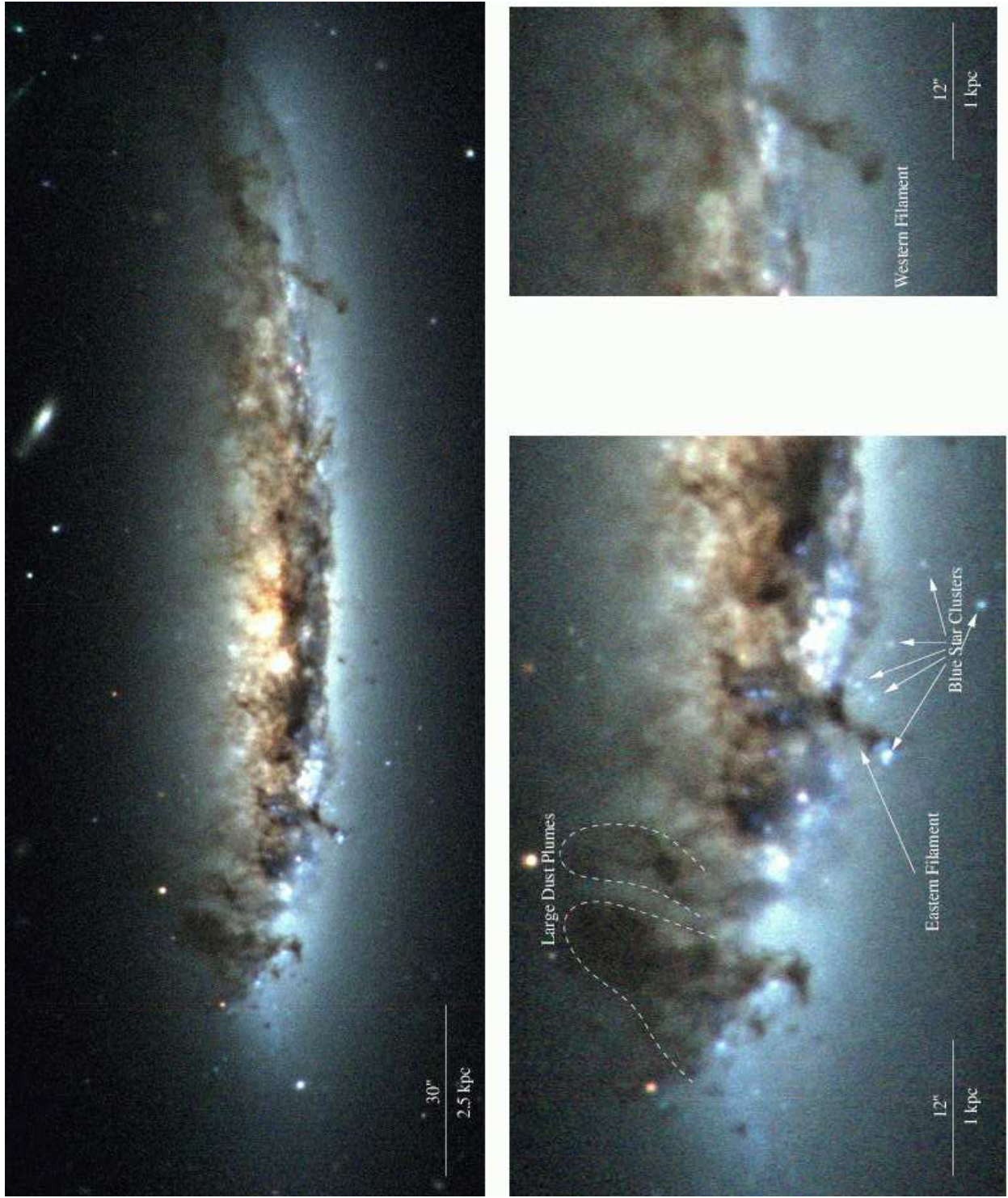}
\caption{BVR color image (left) of NGC~4402, with detail images of the western
  filament (top right) and eastern filament and blue star clusters
  (bottom right).}
\end{figure}

\end{document}